\documentclass[aps,preprint,notitlepage]{revtex4-2}

\usepackage{graphicx}
\usepackage{amssymb}
\usepackage{amsmath}
\usepackage{color}
\usepackage{gensymb}
\usepackage{tikz}
\usepackage{float}
\usepackage{datetime2}

\begin{document}
	\raggedbottom
	
	\title{Construction of isoscattering non-isospectral quantum graphs using the concept of a germ graph}
	
	\author{Omer Farooq}
	\author{Micha{\l} {\L}awniczak}
	\author{Leszek Sirko}
	
	\affiliation{Institute of Physics, Polish Academy of Sciences, Aleja Lotnik\'ow 32/46, 02-668 Warszawa, Poland}
	
\date{\today\quad \DTMcurrenttime}
\begin{abstract}
	
The recently developed concept of germ graphs (P. Kurasov {\it et al.}, Phys. Rev. Research {\bf 7}, L022071 (2025)) is employed to construct families of isocattering, non-isospectral graphs that exhibit dissipation.
More specifically, a new pair of isoscattering non-isospectral graphs is constructed by connecting the existing isoscattering non-isospectral graphs, so called attachments,  to the different contact vertices of the germ graph.
The existing isoscattering non-isospectral graphs contained a two-loop graph labelled $\Gamma_{A2}$ and a three-loop graph labelled $\Gamma_{A3}$. These graphs have an equal total metric length, but different topologies.    The resulting graphs $\Gamma_3$ and $\Gamma_4$ are verified as non-isospectral using their Euler characteristics, while their isoscattering properties are investigated by calculating the one-port scattering matrices numerically. 

	\noindent\textbf{Keywords:} quantum graphs, germ graph, isoscattering, Euler characteristics, scattering matrix
\end{abstract}
	
	\maketitle
	
\section{Introduction}

A central problem in inverse spectral and scattering theory is to determine the underlying structure of a system from its spectral or scattering data \cite{Kac1966,Kurasov2005,Kurasov2024}. In general, such data do not provide a unique reconstruction of the system \cite{Gutkin2001,Kurasov2005,Band2010}. Distinct structures may possess identical eigenvalue spectra, whereas systems with different spectra may generate the same scattering matrix \cite{Band2010,Hul2012,Pistol2021,Lawni2025}. This non-uniqueness demonstrates that neither spectral information nor scattering response alone necessarily determines the topology and metric structure of the underlying system \cite{Gutkin2001,Kurasov2005,Farooq2025,Lawni2025}.

Quantum graphs provide a convenient framework for studying such problems \cite{Kottos1997,Gnutzmann2006,Berkolaiko2013,Kurasov2024}. Their topology and metric structure can be specified independently, while their spectral and scattering properties can be investigated analytically and numerically \cite{Kottos1997,Gnutzmann2006,Pluhar2013,Berkolaiko2013,Kurasov2008Schrodinger,Kurasov2024}. Moreover, quantum graphs can be experimentally realized using microwave networks, which allow their properties to be investigated under controlled experimental conditions \cite{Hul2004,Lawniczak2008,Hul2009,Lawniczak2010,Lawniczak2012,Bialous2016,Rehemanjiang2016,Dietz2017,Lu2020}. More recent studies have extended these  investigations to a broad range of spectral and scattering phenomena \cite{Lawniczak2019SciRep,Lawniczak2019,Chen2020,Chen2021,Farooq2024,lawni2023Acta,lawni2023PRE,Lawni2026}.

Two closed quantum graphs are isospectral when their eigenvalue spectra, including multiplicities, coincide, whereas two open graphs are isoscattering when their corresponding scattering matrices are identical \cite{Band2010,Hul2012,Lawniczak2014,Lawniczak2021}. These two properties are not equivalent. In particular, quantum graphs with different spectra may possess identical scattering matrices \cite{Farooq2025,Lawni2025}.

The relation between spectral and scattering properties of the graphs can be studied using the Titchmarsh--Weyl $M$-function \cite{Kurasov2024}. The $M$-function characterizes a graph at selected contact vertices and is directly related to its scattering properties. In Ref.~\cite{Kurasov2025}, the germ-graph approach was introduced using distinct contact vertices with identical $M$-functions. By attaching the same graph at the corresponding contact vertices, families of isospectral and isoscattering quantum graphs were constructed.

In the present work, we examine whether the germ-graph approach can be extended beyond constructions employing identical attachments. Starting from the germ graph $\Gamma_1$ with two distinguished contact vertices, different graph structures are introduced at the corresponding vertices. The isospectral and isoscattering properties of the attachment graphs and the resulting extended graphs are then examined. The aim is to determine whether isoscattering can persist under such a modification of the germ-graph construction.
\section{Germ Graph and $M$-Function Formalism}

The present construction employs the germ graph introduced in Ref.~\cite{Kurasov2025}. To characterize the graph from a selected contact vertex, we use the Titchmarsh--Weyl  $M$-function. For the single-contact-vertex case considered here, the scalar $M$-function is written as \cite{Kurasov2024}

\begin{equation}
	M_{\Gamma}(k)=\frac{\partial u(v)}{u(v)},
	\label{eq:M_function}
\end{equation}

where $u(v)$ denotes the wave-function value at the contact vertex $v$, and $\partial u(v)$ is the sum of the derivatives directed outward from this vertex.

For a graph coupled to one external channel, its scattering amplitude can be expressed in terms of the corresponding $M$-function as

\begin{equation}
	S_{\Gamma}(k)
	=
	\frac{ik-M_{\Gamma}(k)}
	{ik+M_{\Gamma}(k)}.
	\label{eq:S_M_relation}
\end{equation}

\begin{figure}[H]
	\centering
	\includegraphics[width=0.55\columnwidth]{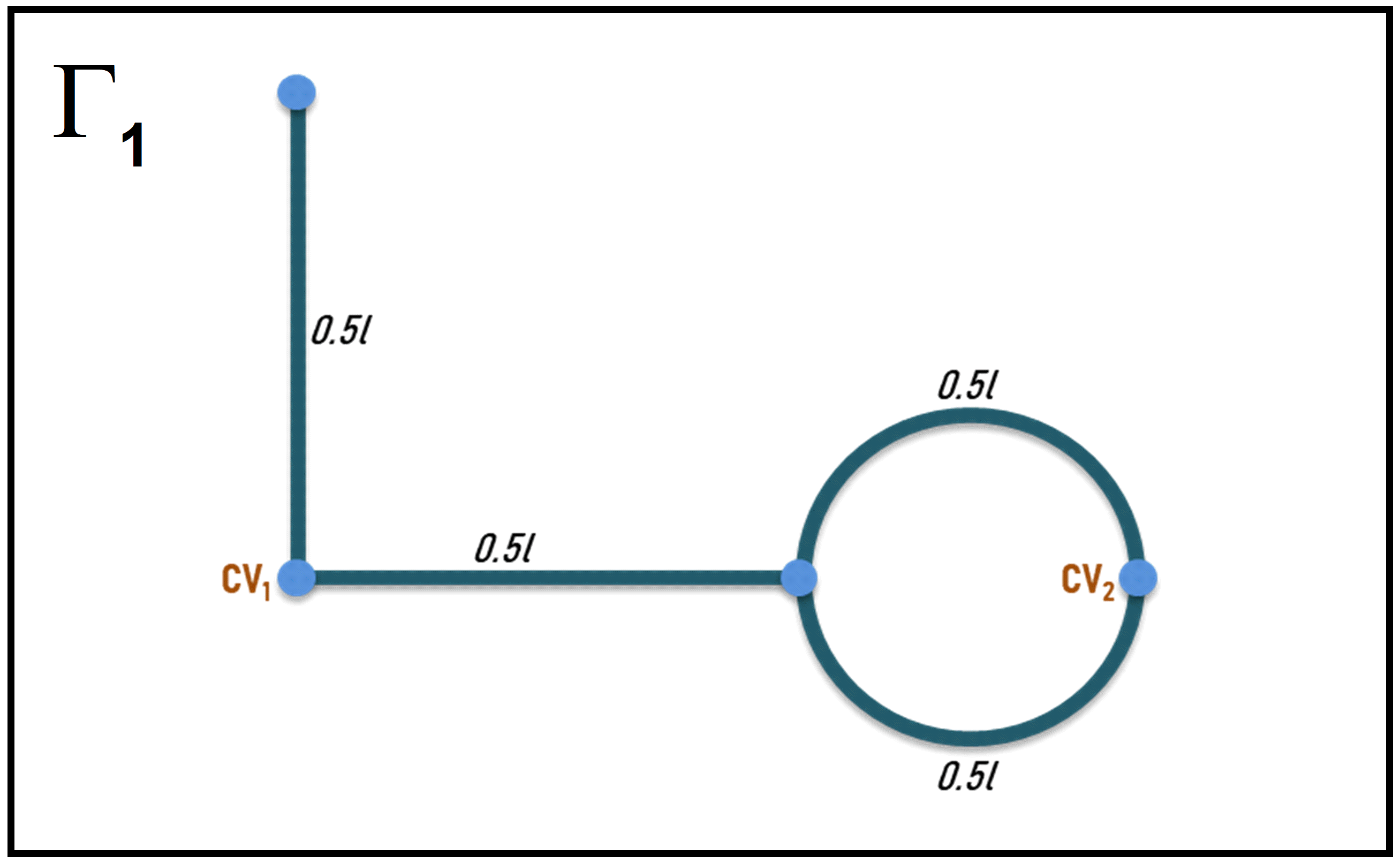}
\caption{The germ graph $\Gamma_1$ with the contact vertices $CV_1$ and $CV_2$.}
	\label{fig:germ_graph}
\end{figure}

The germ graph $\Gamma_1$ used here possesses two distinguished contact vertices, $CV_1$ and $CV_2$. Standard Kirchhoff--Neumann conditions are imposed on the vertices of the compact graph. Its geometry and bond lengths are presented in Fig.~\ref{fig:germ_graph}. Throughout this work, the length scale is set to $l=1~\mathrm{m}$. The total metric length of the germ graph is $L_{\Gamma_1}=2l=2~\mathrm{m}$.

A property of the germ graph established in Ref.~\cite{Kurasov2025} is the equality of the $M$-functions corresponding to the two contact vertices,

\begin{equation}
	M_{CV_1}(k)=M_{CV_2}(k).
	\label{eq:germ_M}
\end{equation}

Thus, the two contact vertices are equivalent with respect to the $M$-function despite their different positions in the graph. This property motivates the construction considered here.
In the present construction, topologically different attachment graphs are associated with the two contact vertices, $CV_1$ and $CV_2$. Their topological and  properties are considered in the following section.


\section{Two-Loop and Three-Loop Attachment Graphs}

\begin{figure}[H]
	\centering
	\includegraphics[width=0.85\columnwidth]{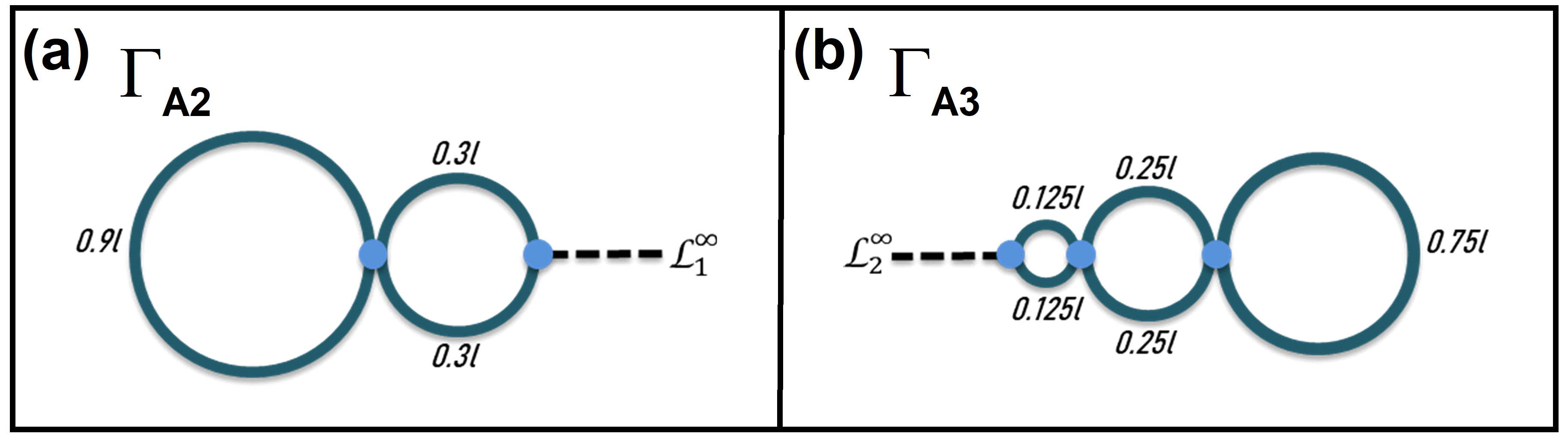}
	\caption{(a) The two-loop graph $\Gamma_{A2}$ and (b) the three-loop graph $\Gamma_{A3}$.}
	\label{fig:attachment_graphs}
\end{figure}

We consider two topologically different attachment graphs, $\Gamma_{A2}$ and $\Gamma_{A3}$, shown in Fig.~\ref{fig:attachment_graphs}. The graph $\Gamma_{A2}$ contains two loops, whereas $\Gamma_{A3}$ contains three loops. Their bond lengths are chosen so that both graphs have the same total metric length,

\begin{equation}
	L_{A2}=2(0.3l)+0.9l
	=
	L_{A3}=2(0.125l)+2(0.25l)+0.75l
	=
	1.5l=1.5~\mathrm{m}.
\end{equation}

\subsection{Isospectrality of the Attachment Graphs}
\label{Non-Isospectrality of the Attachment Graphs}

To examine whether the attachment graphs can be isospectral, we use the Euler characteristic as a simple topological criterion \cite{Lawni2025}. Although the attachment graphs have the same total metric length, their topologies are different. The Euler characteristic of a finite graph is defined as

\begin{equation}
	\chi(\Gamma)=V-E,
\end{equation}

where $V$ and $E$ denote the numbers of vertices and edges, respectively \cite{Kurasov2008Graph,Lawniczak2020Euler,Lawniczak2021Euler,Farooq2023Euler}. For the two attachment graphs,

\begin{equation}
	\chi(\Gamma_{A2})=2-3=-1,
	\qquad
	\chi(\Gamma_{A3})=3-5=-2.
	\label{eq:attachment_Euler}
\end{equation}

Since the Euler characteristic is a spectral invariant of the considered quantum graphs \cite{Kurasov2008Graph,Lawniczak2020Euler,Farooq2023Euler}, the inequality

\begin{equation}
	\chi(\Gamma_{A2})\neq\chi(\Gamma_{A3})
\end{equation}

implies that the two graphs cannot be isospectral. Therefore, their spectra 

\begin{equation}
	\sigma(\Gamma_{A2})\neq\sigma(\Gamma_{A3}).
	\label{eq:attachment_spectra}
\end{equation}
\subsection{Isoscattering Properties of the Attachment Graphs}
\label{sec:isoscattering_attachments}

We next investigate whether the non-isospectral attachment graphs $\Gamma_{A2}$ and $\Gamma_{A3}$ exhibit the same one-port scattering response. The calculation is performed in the directed-bond representation \cite{Kottos1999,Band2012}, in which each physical bond is represented by its two propagation directions. Thus, a graph containing $B$ physical bonds is described in a $2B$-dimensional internal space.

For standard Kirchhoff--Neumann vertex conditions \cite{Kottos1999,Berkolaiko2013, Farooq2022Euler, Farooq2022EdgeEuler}, the scattering amplitudes at a vertex $v_i$ of valency $d_i$ are given by

\begin{equation}
	{}^{N}\sigma_{e,e'}^{(i)}
	=
	\frac{2}{d_i}-\delta_{e,e'},
	\label{eq:vertex_scattering_attachment}
\end{equation}

where $e$ and $e'$ denote the bonds connected to the vertex $v_i$. The corresponding reflection and transmission amplitudes are

\begin{equation}
	r_i=\frac{2}{d_i}-1,
	\qquad
	t_i=\frac{2}{d_i}.
	\label{eq:rt_attachment}
\end{equation}

The complete vertex-scattering matrix is separated into internal and external blocks,

\begin{equation}
	\Sigma=
	\begin{pmatrix}
		S_1 & S_2\\
		S_3 & S_4
	\end{pmatrix},
	\label{eq:blockSigma_attachment}
\end{equation}

where $S_1$ acts within the internal directed-bond space, $S_2$ couples the external channel to the graph, $S_3$ describes the reverse coupling, and $S_4$ gives the direct reflection amplitude in the external channel.

Propagation along the bonds is described by the diagonal matrix

\begin{equation}
	E(\kappa)
	=
	\mathrm{diag}
	\left(
	e^{i\kappa l_1},\ldots,e^{i\kappa l_B},
	e^{i\kappa l_1},\ldots,e^{i\kappa l_B}
	\right),
	\label{eq:E_attachment}
\end{equation}

where $l_j$ denotes the length of the $j$th physical bond. Dissipation (internal absorption) is included through the complex wave number $\kappa=k+i\beta\sqrt{k}$, with $k=2\pi f/c$, where $c=299792458~\mathrm{m/s}$. In the present calculations, $\beta=0.0090~\mathrm{m}^{-1/2}$ is used, consistent with our microwave-network experiments.

Following the corresponding pseudo-orbit formulation \cite{Band2012,Lipovsky2015,Lipovsky2016}, the one-port scattering matrix is calculated as

\begin{equation}
	S(\kappa)
	=
	S_4+
	S_3
	\left[
	I_{2B}-E(\kappa)S_1
	\right]^{-1}
	E(\kappa)S_2.
	\label{eq:S_attachment}
\end{equation}

For each value of the real wave number $k$, the modulus and scattering phase are obtained from the complex scattering amplitude, with

\begin{equation}
	\varphi(\kappa)=-\arg S(\kappa).
	\label{eq:phase_attachment}
\end{equation}

For the attachment graphs, $\Gamma_{A2}$ contains three physical bonds and $\Gamma_{A3}$ contains five, giving internal scattering matrices of dimensions $6\times6$ and $10\times10$, respectively. In both cases, the semi-infinite lead is connected to a vertex of valency $d_i=3$. From Eqs.~\eqref{eq:vertex_scattering_attachment} and \eqref{eq:rt_attachment},

\begin{equation}
	r=-\frac{1}{3},
	\qquad
	t=\frac{2}{3},
\end{equation}

and the corresponding local vertex-scattering matrix is

\begin{equation}
{}^{N}\sigma^{(d_i=3)}
	=
	\begin{pmatrix}
		-\frac{1}{3} & \frac{2}{3} & \frac{2}{3}\\
		\frac{2}{3} & -\frac{1}{3} & \frac{2}{3}\\
		\frac{2}{3} & \frac{2}{3} & -\frac{1}{3}
	\end{pmatrix}.
	\label{eq:d3_matrix_attachment}
\end{equation}

The remaining internal connecting vertices have valency $d_i=4$, for which $r=-1/2$ and $t=1/2$. At the contact vertex, one of the three channels corresponds to the semi-infinite lead and the remaining two to internal directed bonds. Consequently, $S_4=-1/3$, while the nonzero coupling amplitudes entering $S_2$ and $S_3$ are equal to $2/3$. The internal matrix $S_1$ is constructed separately for $\Gamma_{A2}$ and $\Gamma_{A3}$ according to their respective connectivities.

\begin{figure}[H]
	\centering
	\includegraphics[width=0.75\columnwidth]{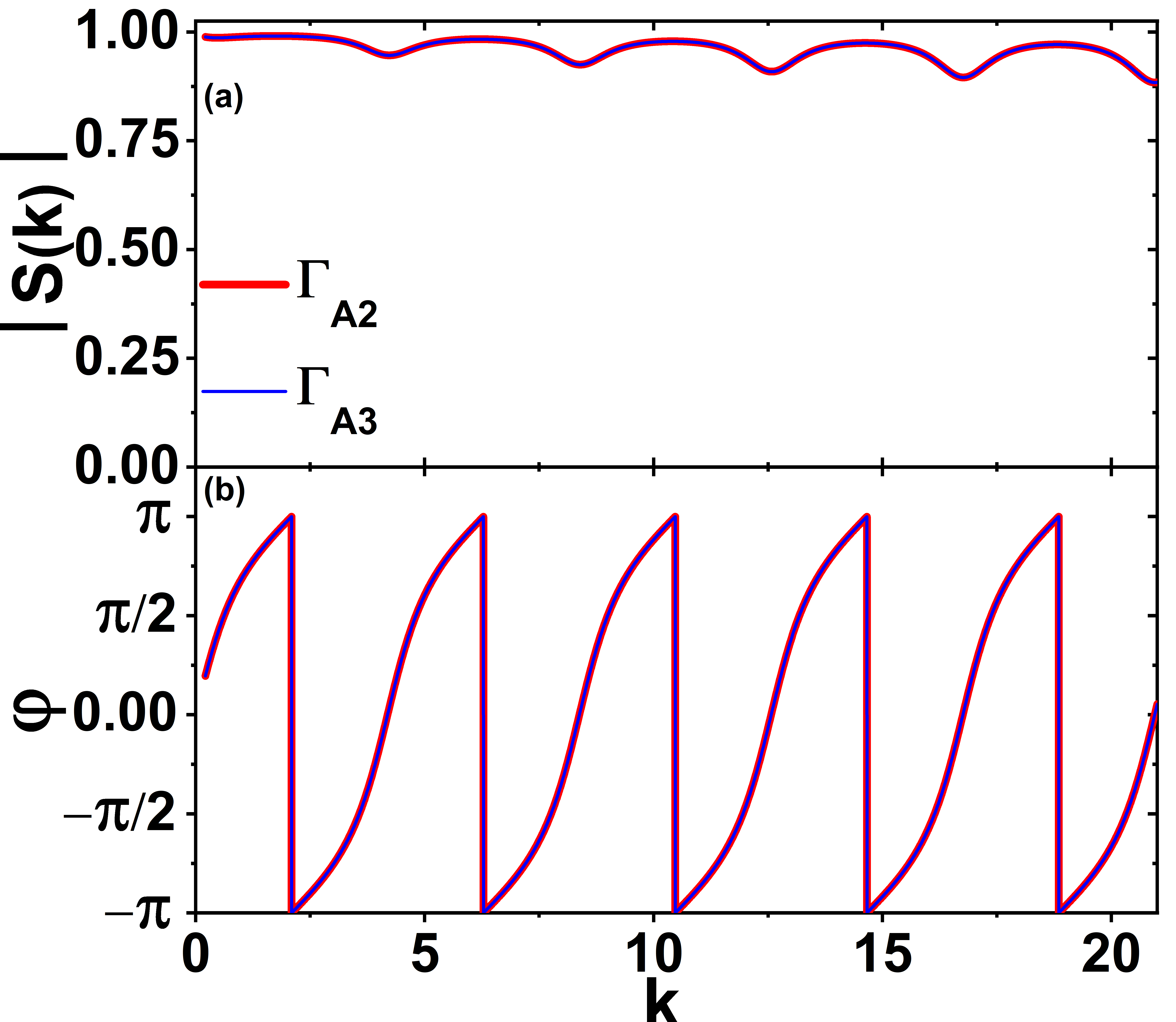}
	\caption{Calculated scattering properties of the attachment graphs
		$\Gamma_{A2}$ (red line) and $\Gamma_{A3}$ (blue line):
		(a) modulus $|S|$ and (b) scattering phase
		$\varphi=-\arg S$ as functions of $k$.}
	\label{fig:scattering_attachment}
\end{figure}

Fig.~\ref{fig:scattering_attachment} shows the calculated modulus and phase of the one-port scattering matrix. The scattering responses of $\Gamma_{A2}$ and $\Gamma_{A3}$ coincide over the investigated wave-number range. Thus,

\begin{equation}
	S_{\Gamma_{A2}}(\kappa)
	\simeq
	S_{\Gamma_{A3}}(\kappa),
	\label{eq:isoscattering_attachment}
\end{equation}
demonstrating that, for the applied absorption strength and associated numerical accuracy,  $|S_{\Gamma_{A2}}(\kappa) - S_{\Gamma_{A3}}(\kappa)| < 10^{-5}$,   $\Gamma_{A2}$ and $\Gamma_{A3}$ are isoscattering and form an isoscattering pair.

\section{Construction and Properties of the Extended Graphs}

\begin{figure}[H]
	\centering
	\includegraphics[width=0.75\columnwidth]{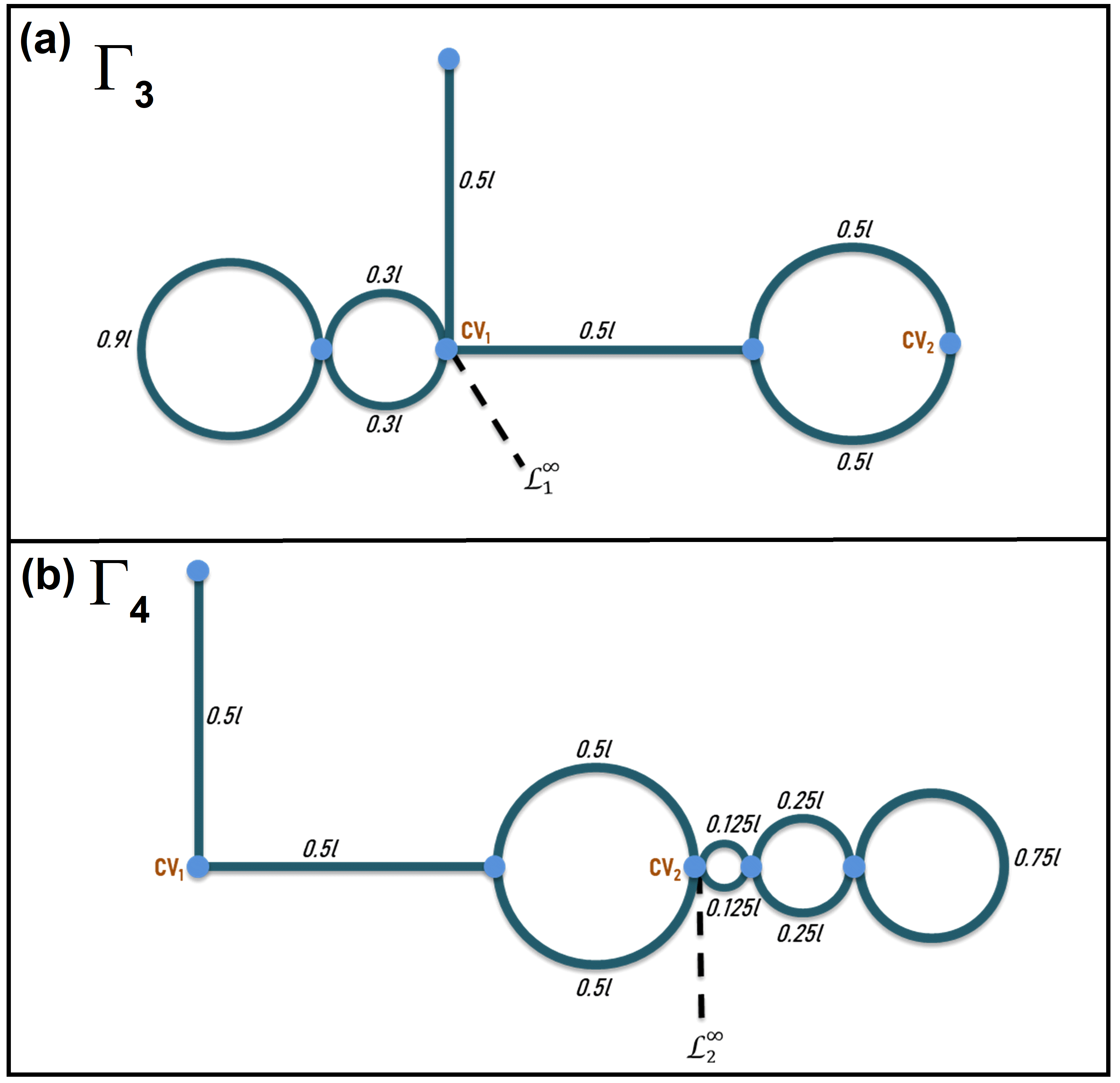}
	\caption{Geometry of the extended graphs: (a) $\Gamma_3$, formed by attaching the  $\Gamma_{A2}$ to the germ graph at $CV_1$, and (b) $\Gamma_4$, formed by attaching the  $\Gamma_{A3}$ at $CV_2$. The semi-infinite leads $\mathcal{L}_1^{\infty}$ and $\mathcal{L}_2^{\infty}$ indicate the corresponding one-port scattering channels.}
	\label{fig:extended_graphs}
\end{figure}

Having established that the two attachment graphs $\Gamma_{A2}$ and $\Gamma_{A3}$ are non-isospectral but isoscattering, we now use them to construct the extended graphs. The two-loop attachment $\Gamma_{A2}$ is connected to the germ graph at $CV_1$ to form $\Gamma_3$, whereas the three-loop attachment $\Gamma_{A3}$ is connected at $CV_2$ to form $\Gamma_4$. The resulting extended graphs are shown in Fig.~\ref{fig:extended_graphs}. For the one-port scattering problem, a semi-infinite lead is attached at the corresponding contact vertex of each graph.

The germ graph has a total metric length $L_{\Gamma_1}=2l$, while both attachments have $L_{A2}=L_{A3}=1.5l$. Hence, the compact parts of the extended graphs have

\begin{equation}
L_{\Gamma_3}=L_{\Gamma_1}+L_{A2}
=
L_{\Gamma_4}=L_{\Gamma_1}+L_{A3}
=
3.5l=3.5~\mathrm{m}.
\end{equation}

Thus, the compact graphs $\Gamma_3$ and $\Gamma_4$ have the same total metric length despite their different topologies.
\subsection{Isospectrality of the Extended Graphs}

The isospectrality of $\Gamma_3$ and $\Gamma_4$ is examined using the Euler characteristic, following the same procedure as in Section~\ref{Non-Isospectrality of the Attachment Graphs}. The semi-infinite leads are not included in the calculation of the Euler characteristic.

The graph $\Gamma_3$ contains five vertices and seven edges, whereas $\Gamma_4$ contains six vertices and nine edges. Therefore,

\begin{equation}
	\chi(\Gamma_3)=5-7=-2,
	\qquad
	\chi(\Gamma_4)=6-9=-3.
	\label{eq:Euler_G3_G4}
\end{equation}

Since the Euler characteristic is a spectral invariant for the considered compact graphs, the inequality

\begin{equation}
	\chi(\Gamma_3)\neq\chi(\Gamma_4)
\end{equation}

implies that the two graphs cannot be isospectral. Consequently,

\begin{equation}
	\sigma(\Gamma_3)\neq\sigma(\Gamma_4).
	\label{eq:spectra_G3_G4}
\end{equation}

\subsection{Isoscattering Properties of the Extended Graphs}

Having established that $\Gamma_3$ and $\Gamma_4$ are non-isospectral, we next examine whether they exhibit the same scattering response. Their one-port scattering matrices are calculated using the same directed-bond procedure introduced in Section~\ref{sec:isoscattering_attachments}.

For $\Gamma_3$, the compact graph contains $B=7$ physical bonds, whereas $\Gamma_4$ contains $B=9$. Thus, their internal directed-bond spaces have dimensions $14$ and $18$, respectively. The vertex-scattering amplitudes are determined by Eqs.~\eqref{eq:vertex_scattering_attachment} and \eqref{eq:rt_attachment}.

For the contact vertex of the extended graphs, the valency is $d_i=5$, giving $r_i=-0.6$ and $t_i=0.4$. The corresponding local vertex-scattering matrix is

\begin{equation}
	{}^{N}\sigma^{(d_i=5)}
	=
	\begin{pmatrix}
		-0.6 & 0.4 & 0.4 & 0.4 & 0.4\\
		0.4 & -0.6 & 0.4 & 0.4 & 0.4\\
		0.4 & 0.4 & -0.6 & 0.4 & 0.4\\
		0.4 & 0.4 & 0.4 & -0.6 & 0.4\\
		0.4 & 0.4 & 0.4 & 0.4 & -0.6
	\end{pmatrix}.
	\label{eq:contact_vertex_matrix}
\end{equation}

The remaining vertices have valencies $d_i=4$ and $d_i=3$, with the corresponding reflection and transmission amplitudes obtained directly from Eq.~\eqref{eq:rt_attachment}.

Using the block decomposition defined in Eq.~\eqref{eq:blockSigma_attachment}, one of the five channels at the contact vertex corresponds to the semi-infinite lead, while the remaining four correspond to internal directed bonds. Therefore,

\begin{equation}
	S_4=-0.6,
\end{equation}

and the nonzero coupling amplitudes entering $S_2$ and $S_3$ are equal to $0.4$. In a suitable ordering of the directed bonds,

\begin{equation}
	S_2=
	\begin{pmatrix}
		0.4\\
		0.4\\
		0.4\\
		0.4\\
		0\\
		\vdots\\
		0
	\end{pmatrix},
	\qquad
	S_3=
	\begin{pmatrix}
		0 & \cdots & 0 &
		0.4 & 0.4 & 0.4 & 0.4 &
		0 & \cdots & 0
	\end{pmatrix}.
	\label{eq:S2S3}
\end{equation}

The internal matrices $S_1$ are constructed separately for $\Gamma_3$ and $\Gamma_4$ according to their respective connectivities. Propagation is described by Eq.~\eqref{eq:E_attachment}, using the corresponding bond lengths of each graph and the same complex wave number and absorption parameter introduced in Section~\ref{sec:isoscattering_attachments}. The one-port scattering matrix is then evaluated from Eq.~\eqref{eq:S_attachment}, and its phase from Eq.~\eqref{eq:phase_attachment}.

\begin{figure}[H]
	\centering
	\includegraphics[width=0.75\columnwidth]{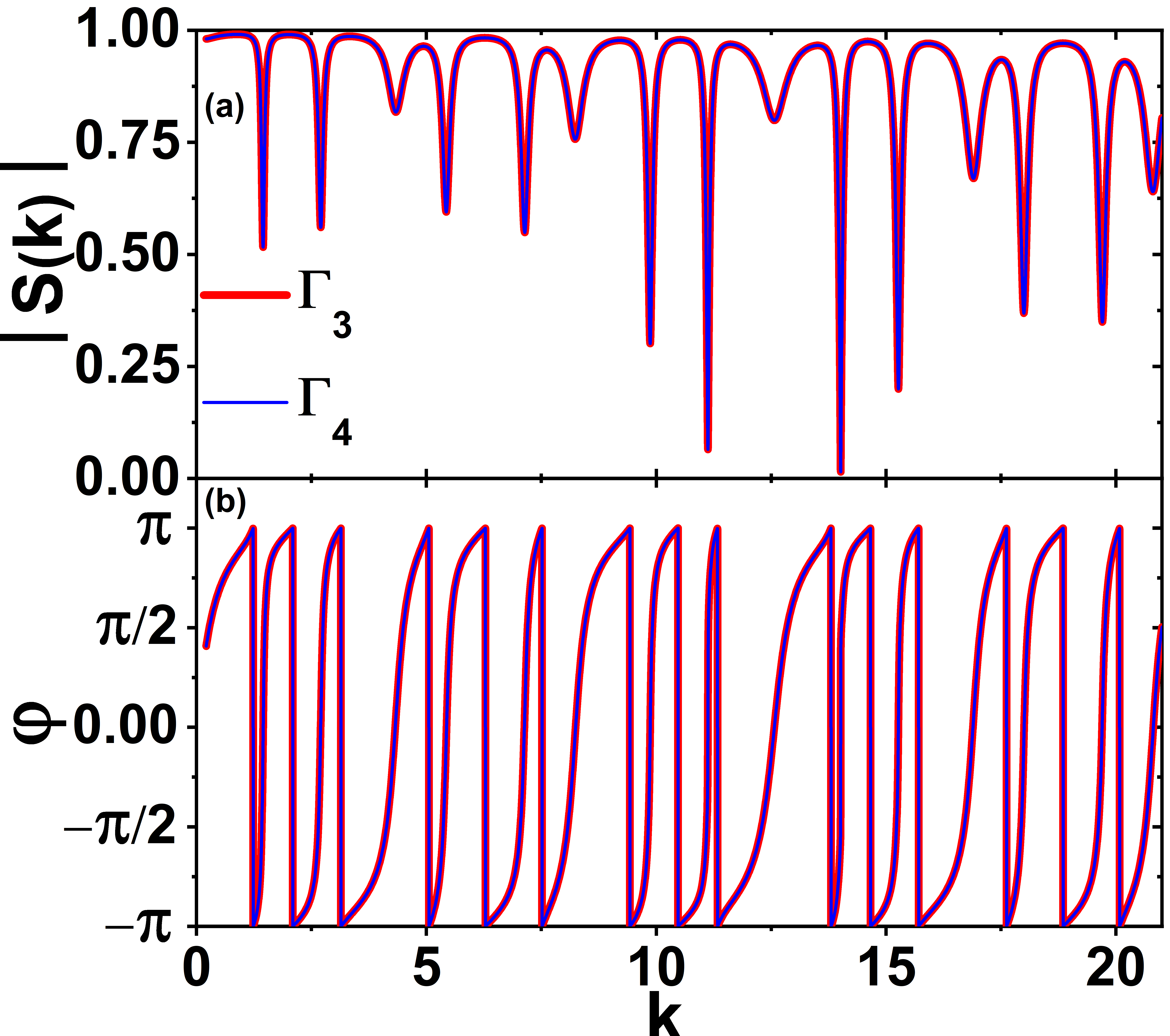}
	\caption{Calculated scattering properties of $\Gamma_3$ (red line) and $\Gamma_4$ (blue line): (a) modulus $|S|$ and (b) scattering phase $\varphi=-\arg S$ as functions of the real wave number $k$. The overlapping curves demonstrate the identical scattering response of the two graphs.}
	\label{fig:scattering}
\end{figure}

Fig.~\ref{fig:scattering} shows the calculated modulus and phase of the one-port scattering matrix. The curves for $\Gamma_3$ and $\Gamma_4$ coincide within numerical accuracy  over the investigated wave-number range. Thus,

\begin{equation}
	S_{\Gamma_3}(\kappa)\simeq S_{\Gamma_4}(\kappa),\\
		\label{eq:isoscattering_numerical}
	\end{equation}
demonstrating that  the graphs $\Gamma_3$ and $\Gamma_4$ are isoscattering.

\section{Summary}

We have shown that the dissipative germ graph can be used to construct the isoscattering, non-isospectral graphs  $\Gamma_3$ and $\Gamma_4$. 
The different Euler characteristics of these graphs prove that they are non-isospectral, even though their numerically calculated one-port scattering responses coincide. Therefore, despite their different internal topologies, the two graphs constitute an isoscattering, non-isospectral pair. This technique can be used to construct other members of the family of graphs that have these properties.
The presented construction of isoscattering non-isospectral graphs  provides a further example of the non-uniqueness of inverse scattering in quantum graphs, demonstrating that distinct graph structures can generate the same scattering response.

\section*{Acknowledgments}

This research was funded by the Institute of Physics of the Polish Academy of Sciences.


	\end{document}